\documentclass[%
 reprint,
superscriptaddress,
 amsmath,amssymb,
 aps,
prb,
]{revtex4-1}

\usepackage{graphicx}
\usepackage{dcolumn}
\usepackage{bm}

\newcommand\ket[1]{\left|#1\right>}
\newcommand{\eq}[1]{\begin{equation}#1\end{equation}}
\newcounter{refc}
\newcommand{\sref}[1]{\addtocounter{refc}{1}\textsuperscript{\arabic{refc}}}
\begin{document}

\title{Topological Hall Effect in Magnetic Topological Insulator Films}

\author{Jian-Xiao Zhang}
\affiliation{%
 Department of Physics, The Pennsylvania State University, University Park, Pennsylvania 16802-6300, USA
}%
\author{Domenico Andreoli}
\affiliation{
 Department of Physics, University of New Hampshire, Durham, New Hampshire 03824, USA
 }
\author{Jiadong Zang}
\affiliation{
 Department of Physics, University of New Hampshire, Durham, New Hampshire 03824, USA
 }
\author{Chao-Xing Liu}%
\affiliation{%
 Department of Physics, The Pennsylvania State University, University Park, Pennsylvania 16802-6300, USA
}%

\date{\today}

\begin{abstract}
Geometric Berry phase can be induced either by spin-orbit coupling, giving rise to the anomalous Hall effect in ferromagnetic materials, or by chiral spin texture, such as skyrmions, leading to the topological Hall effect. Recent experiments have revealed that both phenomena can occur in topological insulator films with magnetic doping, thus providing us with an intriguing platform to study the interplay between these two phenomena. In this work, we numerically study the anomalous Hall and topological Hall effects in a four-band model that can properly describe the quantum well states in the magnetic topological insulator films by combining Landauer-B\"uttiker formula and the iterative Green's function method. Our numerical results suggest that spin-orbit coupling in this model plays a different role in the quantum transport in the clean and disordered limits. In the clean limit, spin-orbit coupling mainly influences the longitudinal transport but does not have much effect on topological Hall conductance. Such behavior is further studied through the analytical calculation of scattering cross-section due to skyrmion within the four-band model. In the disordered limit, the longitudinal transport is determined by disorder scattering and spin-orbit coupling is found to affect strongly the topological Hall conductance. This sharp contrast unveils a dramatic interplay between spin-orbit coupling and disorder effect in topological Hall effect in magnetic topological insulator systems.
\end{abstract}

\maketitle

\section{Introduction}
For electric conductors placed in an external magnetic field, the Lorentz field felt by electrons can lead to a voltage transverse to the electric current, which is known as the Hall effect \cite{hall1879new}.
In magnetic systems, the exchange interaction between electron spin and magnetic moments can give rise to additional topological contributions to the Hall effect.
In a ferromagnetic (FM) system with a strong spin-orbit coupling (SOC), the additional Hall contribution is induced by Berry phases accumulated by the adiabatic motion of quasiparticle on the Fermi surface in the momentum space and normally known as (intrinsic) anomalous Hall effect (AHE) \cite{nagaosa2010anomalous,karplus1954hall}.
On the other hand, when electrons propagate through chiral magnetic structures, e.g. skyrmions, in the real space, they can also feel Berry phase due to the magnetization texture, leading to the so-called ``topological Hall effect (THE)'' (also known as ``geometric Hall effect'') \cite{machida2007unconventional,kanazawa2015discretized,neubauer2009topological,oveshnikov2015berry,ishizuka2018spin}. Both Hall phenomena originate from Berry phase contribution and thus are topological.
An intriguing question is how to understand the topological contribution to the Hall effect
in a magnetic skyrmion system with strong SOC, where the Berry phase exists in both the real and momentum spaces.

Topological insulator (TI) films with magnetic doping, dubbed ``magnetic topological insulator (MTI)'' below,
provide an ideal platform to explore the interplay
between AHE and THE. The coexistence of strong SOC and ferromagnetism in MTI films can result
in a strong AHE \cite{jungwirth2002anomalous}, and the Hall resistance can even achieve the quantized value
when the chemical potential is tuned into the magnetization gap of surface states.
Such phenomenon, known as the quantum anomalous Hall (QAH) effect \cite{yu2010quantized, haldane1988model, liu2008quantum, liu2016quantum}, has been experimentally observed in Cr or V doped (Bi,Sb)$_2$Te$_3$ films \cite{chang2013experimental}.
Furthermore, the surface states in TI film can also mediate Dzyaloshinsky-Moriya (DM) interaction
between magnetic moments due to spin-momentum locking \cite{zhu2011electrically, ye2010spin}.
As a result, chiral magnetic structures, such as skyrmion, are also possible.
Indeed, recent experiments on Cr-doped-(Bi,Sb)$_2$Te$_3$/(Bi,Sb)$_2$Te$_3$ structure \cite{yasuda2016geometric} and Mn-doped Bi$_2$Te$_3$ \cite{liu2017dimensional} have observed a hump in the Hall resistance hysteresis loop at a small magnetic field. The hump structure is attributed to THE while the Hall hysteresis loop implies AHE. Therefore, the interplay between the AHE from ferromagnetism and the THE from magnetic skyrmion will be substantial to understand the electron transport phenomena in MTI films. In addition, MTI is normally highly disordered due to magnetic doping and it is not well understood how the disorder influences the THE in such strong spin-orbit coupled materials.

In this work, we numerically study the magneto-transport of MTI films with a magnetic skyrmion based on a four-band model by combining the iterative Green's function method and the Landauer-Buttiker formalism.
Our numerical results suggest that (1) both AHE and THE can coexist in our model system and the total Hall effect can be decomposed into the summation of these two effects;
(2) in the clean limit, the topological Hall conductance (THC) almost remains constant but the topological Hall resistance (THR) can increase due to the reduction of longitudinal conductance when the SOC is increasing; (3) in the disorder limit, both the THC and THR are increasing with increasing SOC, while longitudinal conductance is not influenced much by SOC. In addition to numerical simulations, we also studied the scattering cross-section of a skyrmion texture analytically with the second-order Born approximation to provide a more theoretical understanding of this system.
Our results are organized as the following. In Sec.\;II, we will describe our model Hamiltonian for the quantum well states in MTI films. In Sec.\;III, we will give our numerical results based on Landauer-Buttiker formalism for the model Hamiltonian and present the corresponding theoretical analysis. The calculation of scattering cross-section will be performed in Sec.\;IV to provide the additional theoretical understanding of the asymmetric scattering for our model Hamiltonian.
The disorder effect is numerically calculated and discussed in Sec.\;V. The conclusion will be drawn in Sec.\;VI.

\section{ Model Hamiltonian}
The TI films can be modeled by a 3D four-band model  in a quantum well (QW)
with an infinite potential along the $z$ direction \cite{zhang2009topological,liu2010model}.
The confinement effect along the $z$ direction can be approximated by choosing $\langle k_z\rangle=0$
and $\langle k_z^2\rangle=(n_b \pi /d)^2$, where $n_b$ is an integer to label the sub-band index and $d$ is the width of the QW \cite{liu2010oscillatory}.
As shown in the Appendix\;A \cite{appendix}, we project the 3D four-band model into the subspace spanned by these QW sub-bands and obtain a 2D four-band BHZ-like model given by
\eq{
H_0(\bm{k}) = M(\bm{k})\sigma_0\tau_z + \alpha(p_x \sigma_x + p_y \sigma_y)\tau_x,
\label{eq:h0}
}
on the basis $(\ket{+ \uparrow},\ket{+ \downarrow}, \ket{- \uparrow}, \ket{- \downarrow})$,
where $\sigma$ and $\tau$ are Pauli matrices for spin and orbital subspaces.
$M(\bm{k})=M_0+Bk_\perp^2$ and $\alpha$ labels the SOC strength.
The parameter $M_0=M+B_1 (n_b \pi /d)^2$ (See appendix for details) depends on the integer sub-band index $n_b$. Depending on the sub-band index $n_b$ (assuming $M<0$ and $B_1>0$), the four band model for the QW sub-bands can be in the inverted regime if $M_0 B<0$ or in the normal regime $M_0 B>0$.
It should be mentioned that the Hamiltonian (\ref{eq:h0}) is block-diagonal with one block set by the basis $(\ket{+ \uparrow}, \ket{- \downarrow})$ and the other block by the basis $(\ket{+ \downarrow}, \ket{- \uparrow})$. These two blocks are related to each other by time-reversal symmetry and are degenerate. This degeneracy will be broken when introducing ferromagnetism or magnetic skyrmion into the system. Since we focus on the transport regime dominated by these QW states in this work,
we expect that multiple QW sub-bands with different $n_b$ will be present at the Fermi energy.
To simplify the problem, we treat QW sub-bands in the Hamiltonian (Eq.\;\ref{eq:h0})
with different $n_b$ independently. Thus, we may choose $M_0$ as an independent parameter and discuss below
the transport behaviors for the parameter $M_0$ in different regimes.
The SOC term ($\alpha$ term) couples the state $\ket{+ \uparrow}$ $(\ket{+ \downarrow})$ to the state $\ket{- \downarrow}$ $(\ket{- \uparrow})$ in different orbital basis, which is different from the conventional Rashba SOC where the SOC term couples different spin states in the same orbital basis.
Here we adopted the Hamiltonian form (Eq.\;\ref{eq:h0}) in Ref.\;\onlinecite{zhang2009topological}, which is equivalent to the more standard Hamiltonian form given in Ref.\;\onlinecite{liu2010model} up to a unitary transformation $U=\textrm{Diag} (1,1,-i,i)$.

\begin{figure}[ht]
\centering
\includegraphics[width=0.35\textwidth]{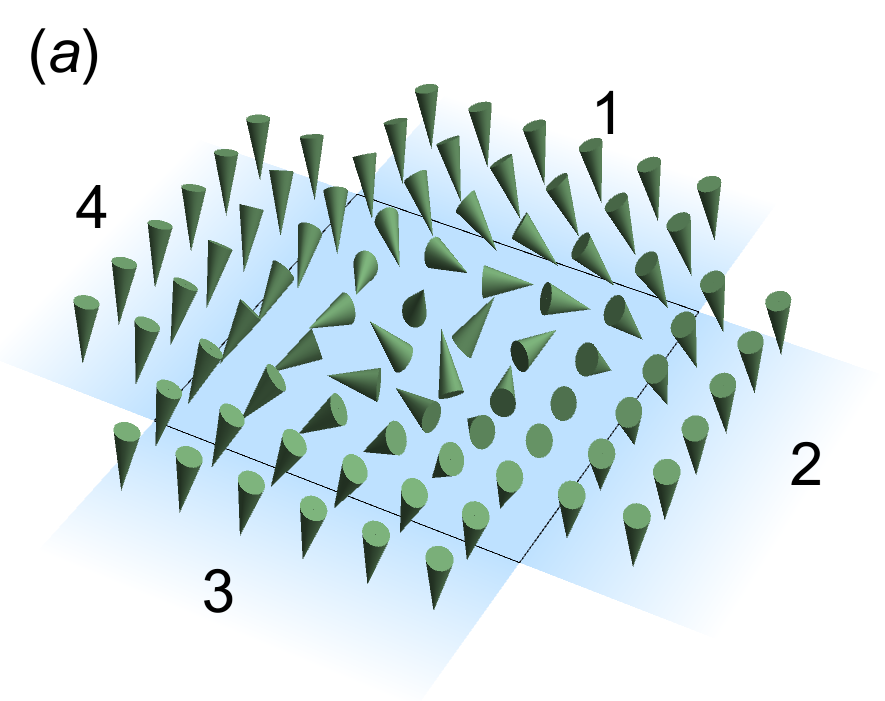}
\includegraphics[width=0.23\textwidth]{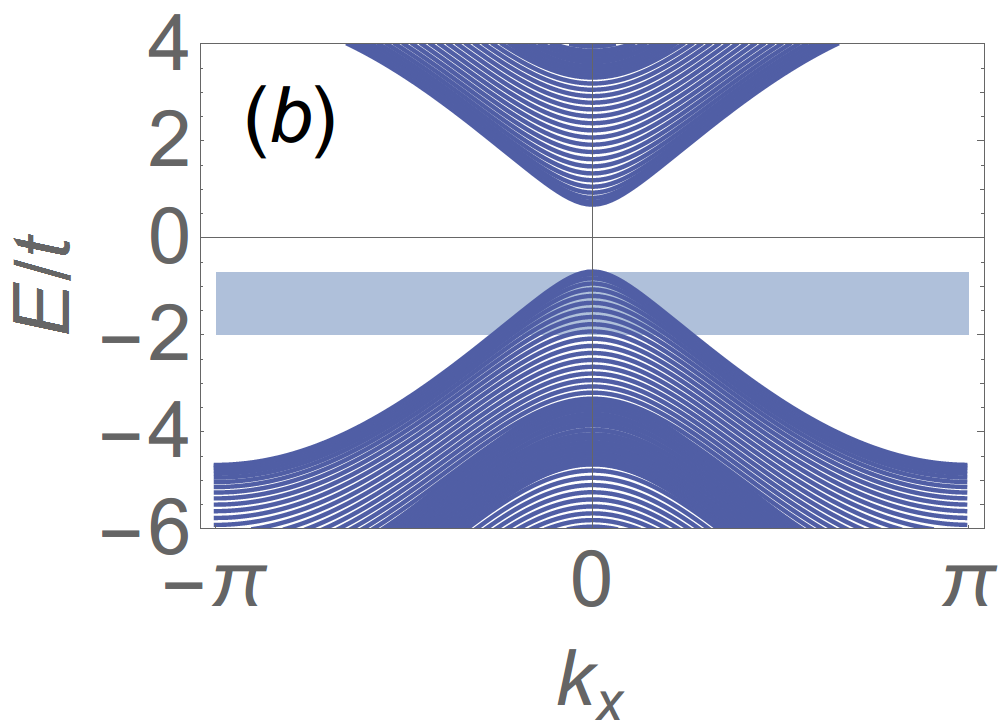}
\includegraphics[width=0.23\textwidth]{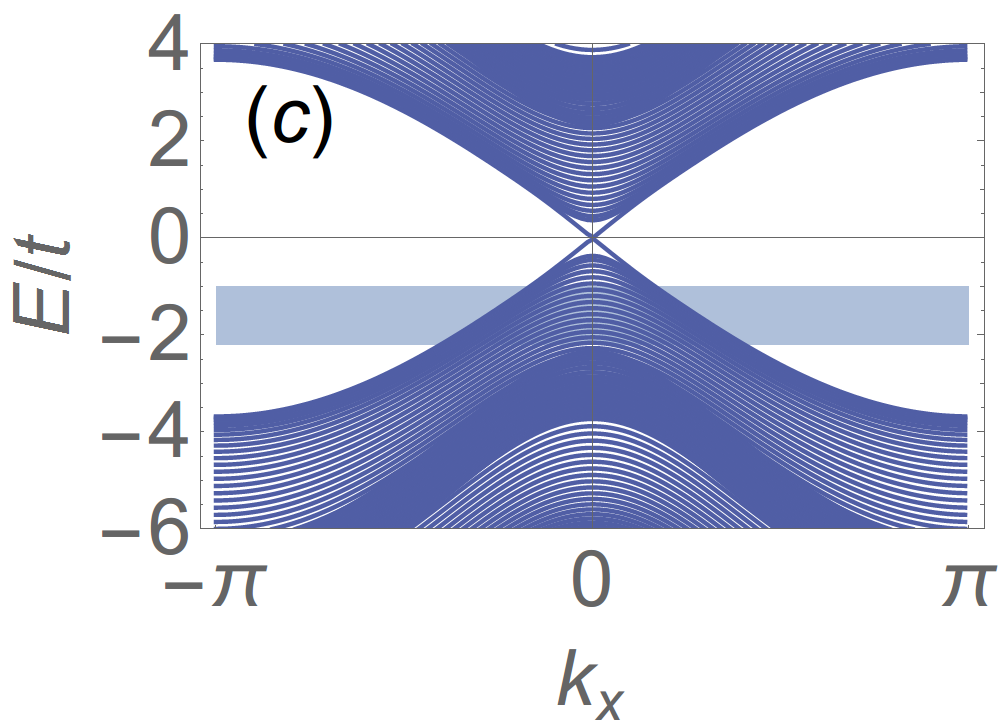}
\caption{(a) Schematic configuration of the system.
The square region in the center represents the sample, and the extended transparent edges represent semi-infinite FM leads. A skyrmion of $m=1, \eta=\pi/2, R = 0.4 L$ is shown at the origin. Cones point to the local magnetic moment direction.
(b, c) Band dispersions for the Hamiltonian Eq.\;\ref{eq:hamiltonian} for two parameter sets (i) and (ii) (see main text).
A periodic boundary condition is applied on the $y$ direction. The shaded regions show the energy range for the transport calculations in Fig.\;\ref{fig:the01m1} to \ref{fig:THEenhancement}.
}
\label{fig:config}
\end{figure}

For MTI, the exchange interaction between electron spin and magnetic moment is given by
\eq{
H_{\textrm{ex}}(x,y) = \bm{m} (x, y) \cdot \bm{\sigma} \tau_0,
\label{eq:Hex}
}
where $\bm{m}(x,y)$ represents the magnetization.
The magnetic skyrmion texture can be taken into account by choosing the $\bm{m}$ configuration
\eq{
\begin{split}
\bm m &= m_0(\sin \theta  \cos \phi ,\sin \theta  \sin \phi
   ,\cos \theta )\\
\theta &= \pi  \tanh \left(\frac{\rho }{R}\right) \\
\phi &= n \varphi + \eta,
\label{eq:skyrmionmag}
\end{split}
}
where $m_0$ represents the magnetization strength, $\theta$ and $\phi$ label the magnetization direction,
$\rho$ and $\varphi$ define the spatial polar coordinates with $(x,y)=(\rho\cos\varphi,\rho\sin\varphi)$,
and $R$ is the radius of the skyrmion. The chirality of the skyrmion is characterized by the integer number $n$,
which is chosen to be $+1$ (a single skyrmion) or $-1$ (a single anti-skyrmion) below.
The parameter $\eta$ denotes the helicity phase, which is an irrelevant parameter.
It should be pointed out that the unitary transformation $U$ should be applied to the Hamiltonian (\ref{eq:Hex}) in order to be consistent with the Hamiltonian (\ref{eq:h0}). However, we find the THE only depends on the chirality of skyrmion texture, which is unchanged under the transformation $U$, and thus we can still use the Hamiltonian (\ref{eq:Hex}) to describe skyrmion texture.
Physically, the magnetic skyrmion can be energetically stabilized by the interplay of the Zeeman coupling and
DM interaction in MTI films \cite{koshibae2016theory}. In this study, we assume the skyrmion structure in our system (Fig.\;\ref{fig:config}(a))
and focus on the influence of skyrmion on magneto-transport.

Due to the absence of translation symmetry in a system with a single skyrmion,
we numerically explore magneto-transport directly in the real space. To perform such calculation, we implement the tight-binding regularization on the Hamiltonian (\ref{eq:h0}) and (\ref{eq:Hex}), which is given by
\eq{
\begin{split}
	\hat{H} &= \hat{H}_0 + \sum_i \Psi_i^\dagger H_{\textrm{ex}} \Psi_i ,\\
\hat{H}_0 &= \sum_i \Psi_{i}^\dagger \epsilon_i   \Psi_{i} + \sum_{\langle i,j\rangle}(V_{ij} \Psi_{j}^\dagger \Psi_{i} + \text{h.c.}),
\label{eq:hamiltonian}
\end{split}
} where $\Psi^\dagger_{i}$ presents $(c^\dagger_{i,+\uparrow}, c^\dagger_{i,+\downarrow}, c^\dagger_{i,-\uparrow}, c^\dagger_{i,-\downarrow})$ at the position $ i = (i_x,i_y)$. $\epsilon_i$ is the on-site energy and $V_{ij}$ is the hopping matrix between the nearest neighbors $i$ and $j$. Both $\epsilon_i$ and $V_{ij}$ are 4 by 4 matrices and their detailed forms can be related to those in the continuous model (Eq.\;\ref{eq:h0}), as listed in the Appendix\;B\cite{appendix}. The consistency between the tight-binding Hamiltonian (\ref{eq:hamiltonian}) and the continuous Hamiltonian (\ref{eq:h0}) and (\ref{eq:Hex}) is also discussed in the Appendix\;B\cite{appendix}.

We consider a 2D square lattice with the side length $L$. The skyrmion texture is located at the center of the lattice,
as shown in Fig.\;\ref{fig:config}(a). 
Four semi-infinite leads with the width $L$, labeled as 1 to 4 in Fig.\;\ref{fig:config}, 
are attached to each side of the square lattice.
We adopt the recursive Green's function method \cite{datta1997electronic} to evaluate the transmission coefficient $T_{pq}$
between the leads $p$ and $q$ ($p,q=1,2,3,4$).
The relationship between currents and voltages is calculated using the Landauer-B\"uttiker formalism
\begin{eqnarray}\label{eq:LB}
I_p=e^2/h \sum_{q\neq p}\left(T_{qp}V_p-T_{pq}V_q\right).
\end{eqnarray}
Due to the charge conservation of the whole system, the matrix $T$ is singular. Without loss of generality, we set $V_4=0$ and remove the corresponding column/row in $T$.
To set up a Hall configuration, we consider a current flow from the lead 1 to 3, as setting $I_1 = -I_3 = I$ and $I_2=0$. Voltages of the leads are calculated through $(V_1,V_2,V_3)^\intercal = \frac{h}{e^2}T^{-1} (I_1,I_2,I_3)^\intercal$, where the matrix $T$ is the $3\times 3$ transmission matrix.
The longitudinal resistance $R_{xx}$ and the Hall resistance $R_{xy}$ can be extracted by $R_{xx} = (V_1-V_3)/I$ and  $R_{xy} = V_2/I$.

\section{Numerical results and Analysis in the Clean limit}

\begin{figure*}[tp]
\centering
\includegraphics[width=0.45\textwidth]{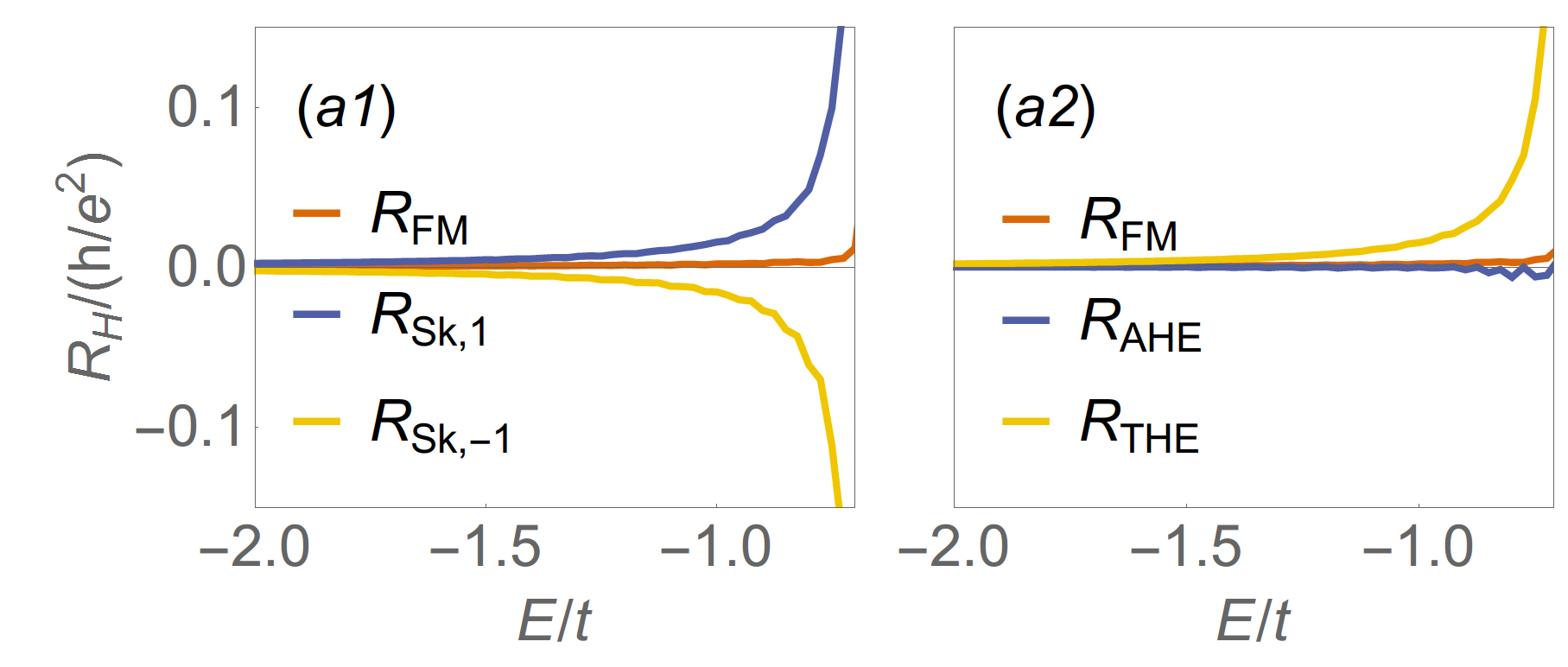}
\includegraphics[width=0.45\textwidth]{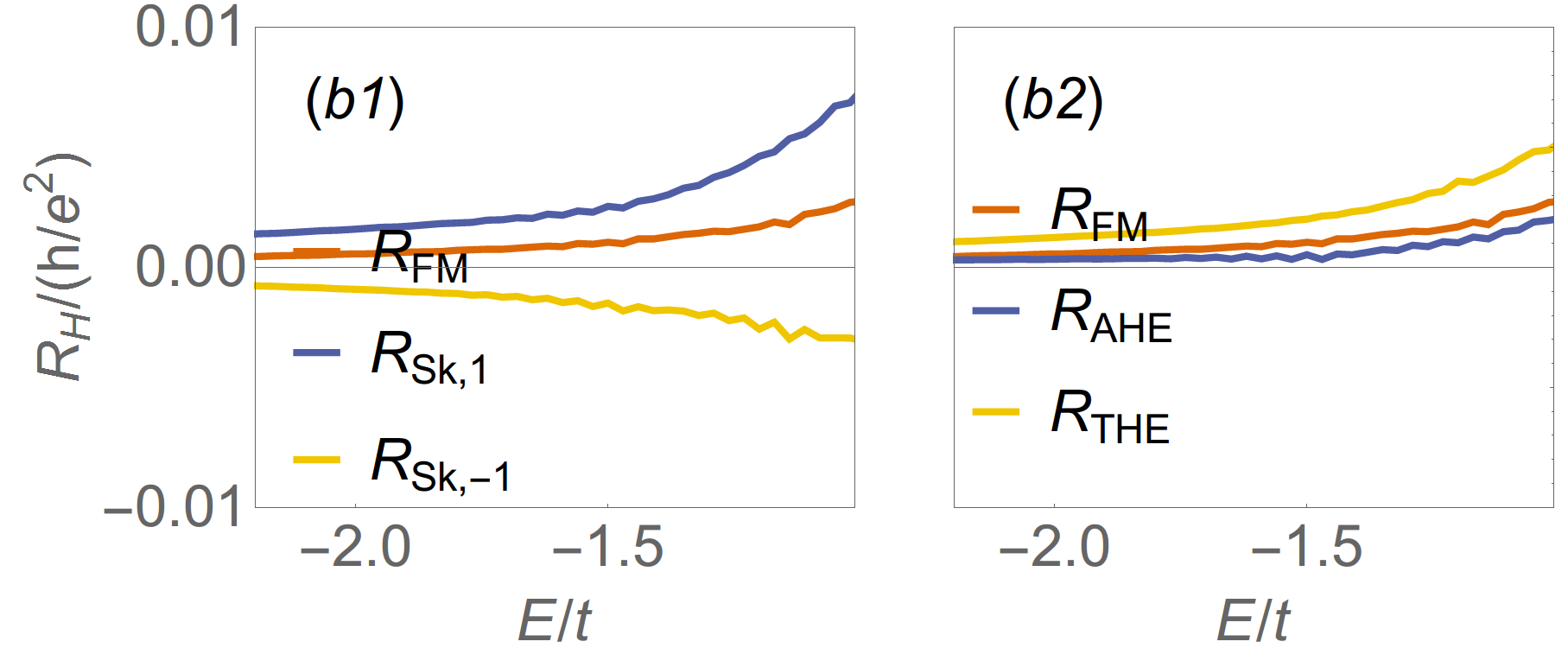}
\caption{(a1, b1) Red, blue and yellow lines represent Hall resistance in FM, skyrmion and anti-skyrmion cases as the function of Fermi energy, for the parameter sets (i) and (ii), respectively.
SOC strength $\alpha/t=2$.
(a2, b2) Blue and yellow lines are the extracted AHE and THE contribution $R_{\textrm{AHE}}$ and $R_{\textrm{THE}}$ from Eq.\;\ref{eq:Hall_skyrmion_decomposition}. Red lines are the same in (a1, b1) representing the FM contribution $R_{\textrm{FH}}$ as a comparison.
Note that $R_{\textrm{AHE}}$ and $R_{\textrm{FH}}$ are close in value in a wide range of energy, suggesting the skyrmion texture has a minimum reduction on the AHE effect.
}
\label{fig:the01m1}
\end{figure*}

\begin{figure*}[tp]
\centering
\includegraphics[width=0.45\textwidth]{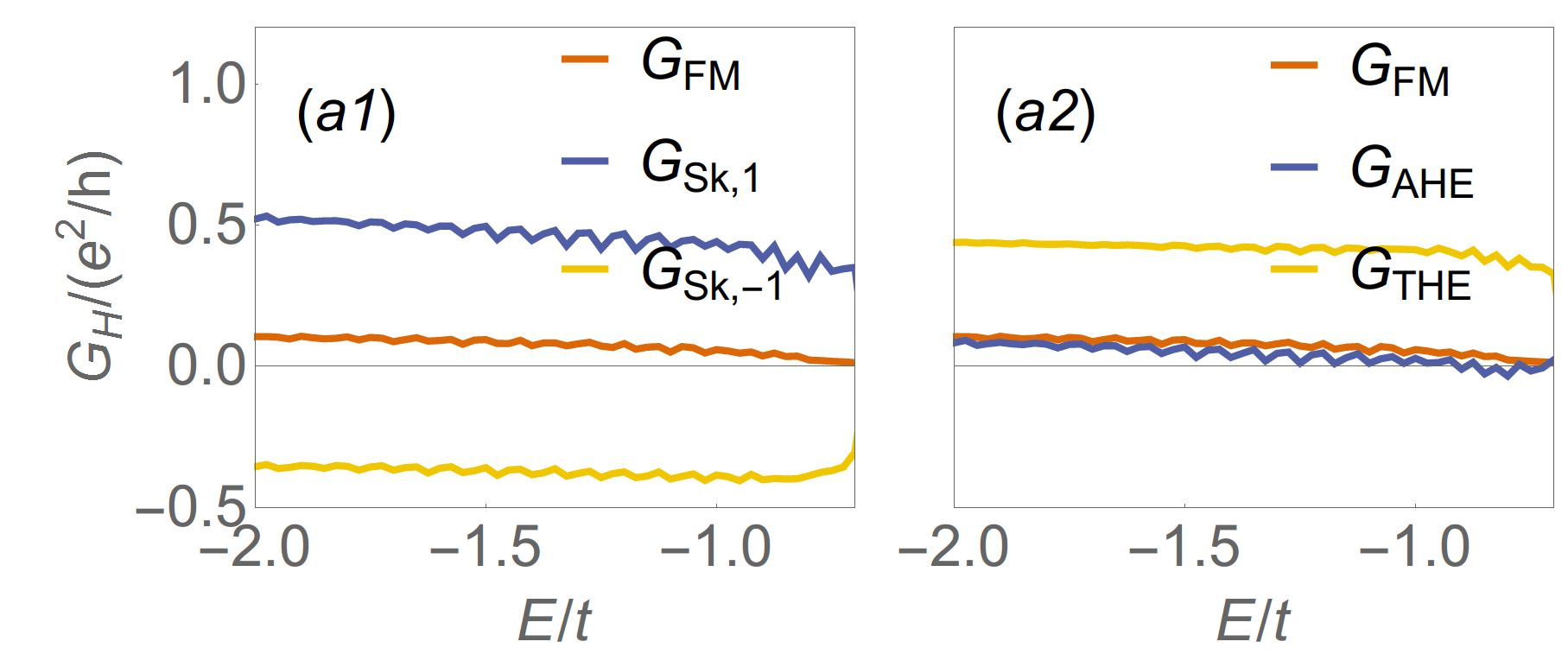}
\includegraphics[width=0.45\textwidth]{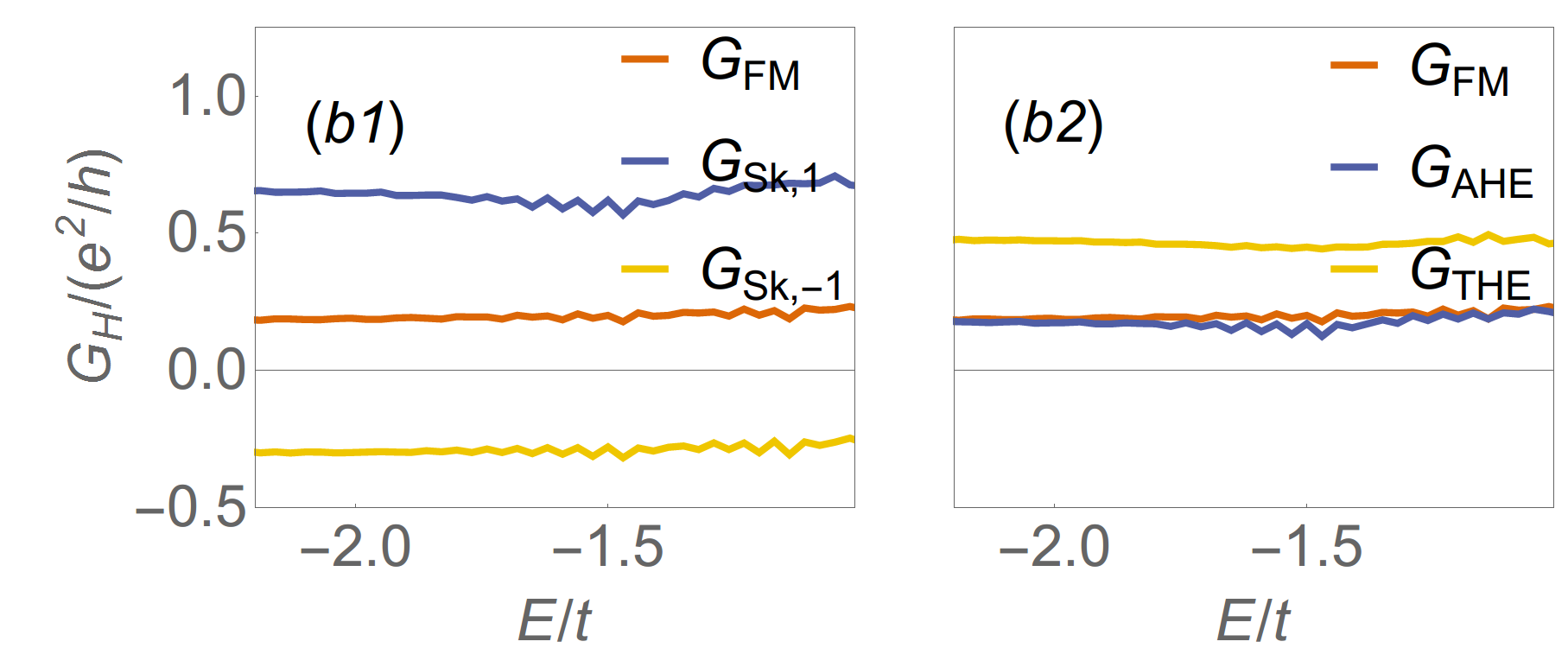}
\caption{The same system as Fig.\;\ref{fig:the01m1} expressed in conductances. (a1, b1) the Hall conductances for the parameter sets (i) and (ii), respectively. (a2, b2) the extracted decomposition into AHE and THE contribution. The ferromagnetic system is marked with red lines. Note that $G_{\textrm{FM}}$ is also close to $G_{\textrm{AHE}}$, showing that the decomposition approximation works for both resistance and conductance.
}
\label{fig:theg01m1}
\end{figure*}

Fig.\;\ref{fig:the01m1}(a1) and (b1) reveal the Hall resistance as a function of the Fermi energy in different parameter regimes. Here we choose a system of length $L =80 a_0$ and Skyrmion radius $R = 0.4 L$. Hopping parameters are expressed in term of the nearest-neighbor hopping strength $t \equiv B/a_0^2$ (See Appendix\;B\cite{appendix}). Here we consider two sets of parameters, one for the trivial regime, denoted as (i), and the other for the QAH regime, denoted as (ii) below. While all the other parameters are the same ($m_0=4/3,\alpha/t=2, a_0=1, B=1$) for the parameter sets (i) and (ii), the parameter $M_0$ is chosen to be different ($M_0 = 2$ for (i) and $M_0=1$ for (ii)). To see the topological property of the full Hamiltonian with these two parameter sets, we may consider the FM case with $\bm{m}(x,y)=m_0(0,0,1)$. For the parameter set (i), we notice that both blocks of bands in the Hamiltonian (\ref{eq:hamiltonian}) are in the normal regime since $(M_0\pm m_0)B>0$. In contrast, the system for the parameter set (ii) is in the QAH regime since one block is in the normal regime $(M_0+m_0)B>0$ while the other is in the inverted regime $(M_0-m_0)B<0$. The corresponding energy dispersions for a slab configuration with these two parameter sets are shown in Fig.\;1(b) and (c), from which one can see a full gap for the parameter set (i) and gapless chiral edge states appear in the bulk gap for the parameter set (ii).
In this work, we focus on the transport behavior of the metallic regime when the Fermi energy $E_f$ crosses one
valence band top ($E_f/t \in [-2, -0.7]$ for the parameter set (i) and $[-2.6,-1]$ for the parameter set (ii) in Fig.\;1(b) and (c)). For the purpose of the quantized conductance within the gap, the parameter set (i) and (ii) represent a comparison between a trivial and a non-trivial gap.
The difference is briefly discussed in Appendix\;D\cite{appendix}.
For the transport calculation, three magnetic configurations, namely ferromagnetism ($\hat m = -m_0 \hat z$), a skyrmion ($n=+1$) and an anti-skyrmion ($n=-1$) are considered and the corresponding Hall resistances $R_{\textrm{FM}}$ , $R_{\textrm{Sk},+1}$ and $R_{\textrm{Sk},-1}$ are shown by the yellow, red and blue lines in Fig.\;\ref{fig:the01m1}(a1) and (b1) for the parameter sets (i) and (ii), respectively. One can clearly see that $R_{\textrm{Sk},+1}$ is much larger than $R_{\textrm{FM}}$, while $R_{\textrm{Sk},-1}$ has the opposite sign.
For the FM case, the Hall resistance
only originates from the AHE, while in the skyrmion cases, both THE and AHE can contribute due to the coexistence of strong SOC and chiral magnetic structure.
We expect the THE (AHE) is dependent (independent) on the chirality of the skyrmions.
Therefore, we can decompose the Hall resistances $R_{\textrm{Sk},\pm 1}$ into chirality dependent part, $R_{\textrm{THE}}$, and independent part, $R_{\textrm{AHE}}$,
\begin{equation}
	R_{\textrm{Sk},n}=R_{\textrm{AHE}}+ n R_{\textrm{THE}},
	\label{eq:Hall_skyrmion_decomposition}
\end{equation}
where the index $n$ stands for the chirality of the skyrmion.

Based on the decomposition of Eq.\;\ref{eq:Hall_skyrmion_decomposition}, Fig.\;\ref{fig:the01m1}\;(a2) and (b2) depict $R_{\textrm{AHE}}$ (blue line) and $R_{\textrm{THE}}$ (yellow line)
as a function of Fermi energy for the parameter sets (i) and (ii), respectively.
In addition, $R_{\textrm{FM}}$ is shown by the red line.
Fig.\;\ref{fig:the01m1}\;(a2) and (b2) show the following features. (1) $R_{\textrm{FM}}$ generally shows a similar behavior as the blue line of $R_{\textrm{AHE}}$ (except that the Fermi energy is close to band gap), and thus the magnetic skyrmion does not have a strong influence on the AHE in the metallic regime and validates the decomposition of the Hall resistance. (2) $R_{\textrm{AHE}}$ is much larger for the parameter set (i) than that for (ii), due to the energy range (shaded area in Fig.\;\ref{fig:config}(b,c)) is closer to band center for parameter set (ii). (3) For both parameter sets, we notice that $R_{\textrm{THE}}$ increases rapidly when the Fermi energy is tuned towards the valence band top.

\begin{figure}[htp]
\centering
\includegraphics[width=0.5\textwidth]{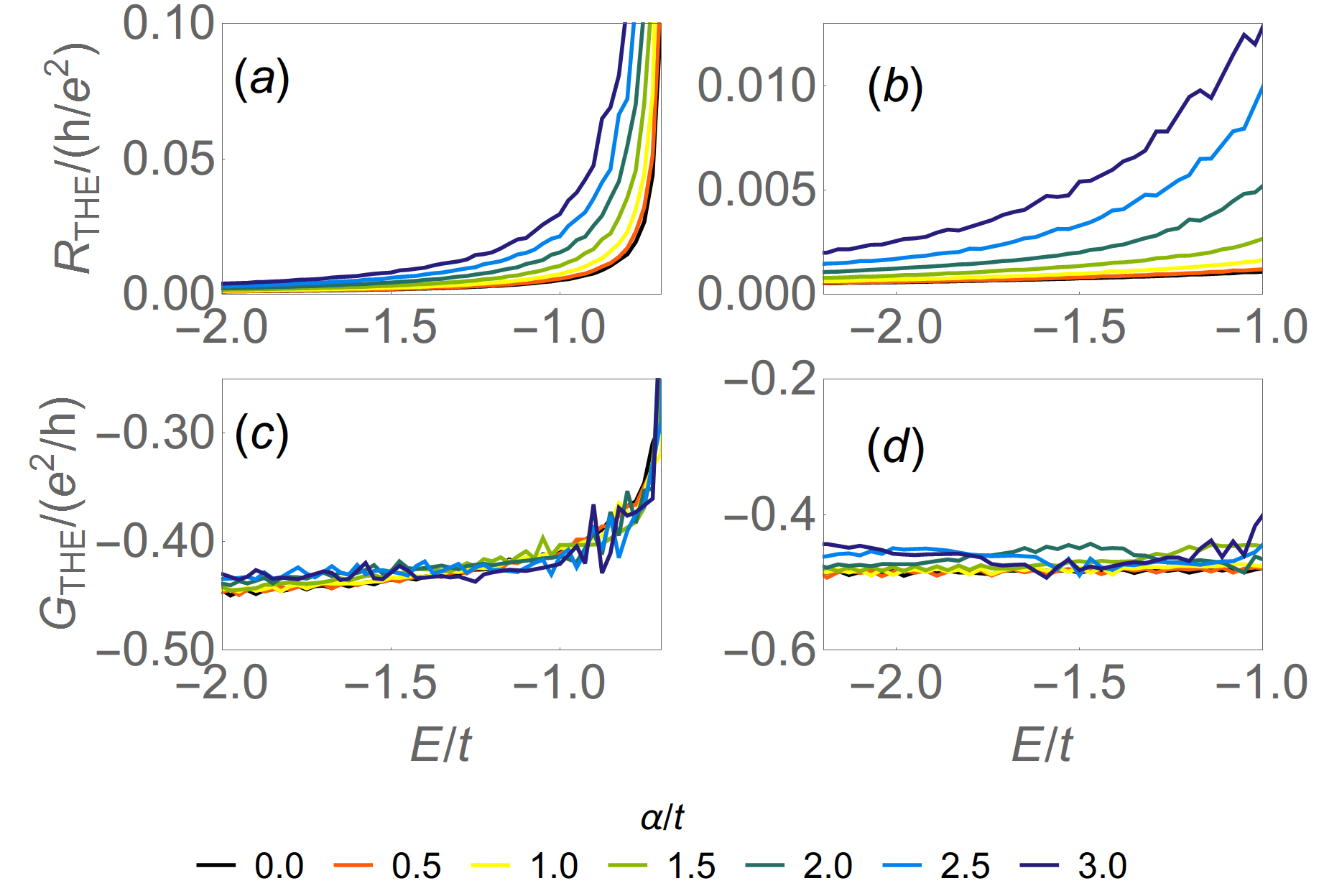}
\caption{(a) and (b) show THR $R_{\textrm{THE}}$ as the function of Fermi energy and SOC strength $\alpha$, for the parameter sets (i) and (ii), respectively. (c) and (d) show THC $G_{\textrm{THE}}$ decomposed in the similar way. }
\label{fig:THEenhancement}
\end{figure}

We next turn to the Hall conductance with the same parameters, as shown in Fig.\;\ref{fig:theg01m1}. Here the red, blue and yellow lines are for the Hall conductance with the FM, skyrmion ($n=+1$) and anti-skyrmion ($n=-1$) configurations in Fig.\;\ref{fig:theg01m1}\;(a1) and (b1) for two parameter sets. A similar decomposition
\begin{equation}
	G_{\textrm{Sk},n}=G_{\textrm{AHE}}+n G_{\textrm{THE}}
	\label{eq:Hall_skyrmion_decomposition_conductance}
\end{equation}
is considered and the corresponding $G_{\textrm{AHE}}$ and $G_{\textrm{THE}}$ are plotted in Fig.\;\ref{fig:theg01m1}\;(a2) and (b2), which show the following features. (1) The decomposition of Hall conductance also remains valid in most energy ranges, as indicated by the coincidence between $G_{\textrm{AHE}}$ and $G_{\textrm{FM}}$ in most energy ranges. (2) In contrast to Hall resistance, the Hall conductance is almost a constant in the whole metallic region for both parameter sets. (3) $G_{\textrm{AHE}}$ (or $G_{\textrm{FM}}$) is smaller for the parameter set (i) compared to that for (ii) while $G_{\textrm{THE}}$ is comparable for both parameter sets.
Fig.\;\ref{fig:THEenhancement}(a) and (b) ((c) and (d)) reveal the THE contribution $R_{\textrm{THE}}$ ($G_{\textrm{THE}}$) from the decomposition of Eq.\;\ref{eq:Hall_skyrmion_decomposition} (Eq.\;\ref{eq:Hall_skyrmion_decomposition_conductance}) as a function of the Fermi energy for different SOC strength $\alpha$ for both parameter sets. An enhancement of $R_{\textrm{THE}}$ is found while $G_{\textrm{THE}}$ remains almost unchanged when increasing the SOC strength or the Fermi energy for both parameter sets.
$G_{\textrm{THE}}$ is only found to drop when the Fermi energy is close to the band gap (insulating regime).


To understand our numerical results, we will next present a theoretical analysis of the transport behavior of the model based on our numerical simulation of the Landauer-Buttiker formalism.
The symmetry property of the transmission matrix $T_{pq}$ in the Eq.\;\ref{eq:LB} will be first analyzed.
For the FM and $n=+1$ skyrmion cases, the system respects the $C_4$ rotation symmetry, while for the anti-skyrmion with $n=-1$, the system possesses the $S_4$ improper rotation symmetry. In all cases, we find the transmission matrix elements can be characterized by three independent parameters based on the following relations
\eq{
\begin{split}
T_{13} = T_{24} = T_{31} = T_{42} \equiv -a \\
T_{14} = T_{21} = T_{32} = T_{43} \equiv -b \\
T_{12} = T_{23} = T_{34} = T_{41} \equiv -c \\
\end{split}
}
where $a$, $b$ and $c$ can be understood as the probability of electronic modes going straight, turning left and turning right, respectively, after they entered the spin-textured structure from any of the leads.
With this simplification, direct calculations from the Landauer-B\"uttiker formalism give rise to the Hall and longitudinal resistance as
\eq{
\begin{split}
R_{xy} &= \frac{(b-c)}{2 a^2+2 a (b+c)+b^2+c^2} \approx \frac{b-c}{2 (a+b)^2} \equiv \frac{\Delta}{2 \beta^2} \\
R_{xx} &= \frac{(2 a+b+c)}{2 a^2+2 a (b+c)+b^2+c^2} \approx \frac{1}{a+b} \equiv \frac{1}{\beta},
\label{eq:resistance_analytical}
\end{split}
}
where we define the parameters $\beta \equiv a+b$ to be the transmission probability and $\Delta \equiv (b-c)$ to be the asymmetric scattering between the left and right directions.
We further assume $\Delta\ll b\approx c$, which can be justified based on our numerical calculations for both parameter sets.
The corresponding Hall and longitudinal conductance is
\eq{
\begin{split}
G_{xy} \approx -\Delta/2 \\
G_{xx} \approx \beta.
\label{eq:conductance_analytical}
\end{split}
}

\begin{figure}[htp]
\centering
\includegraphics[width=0.5\textwidth]{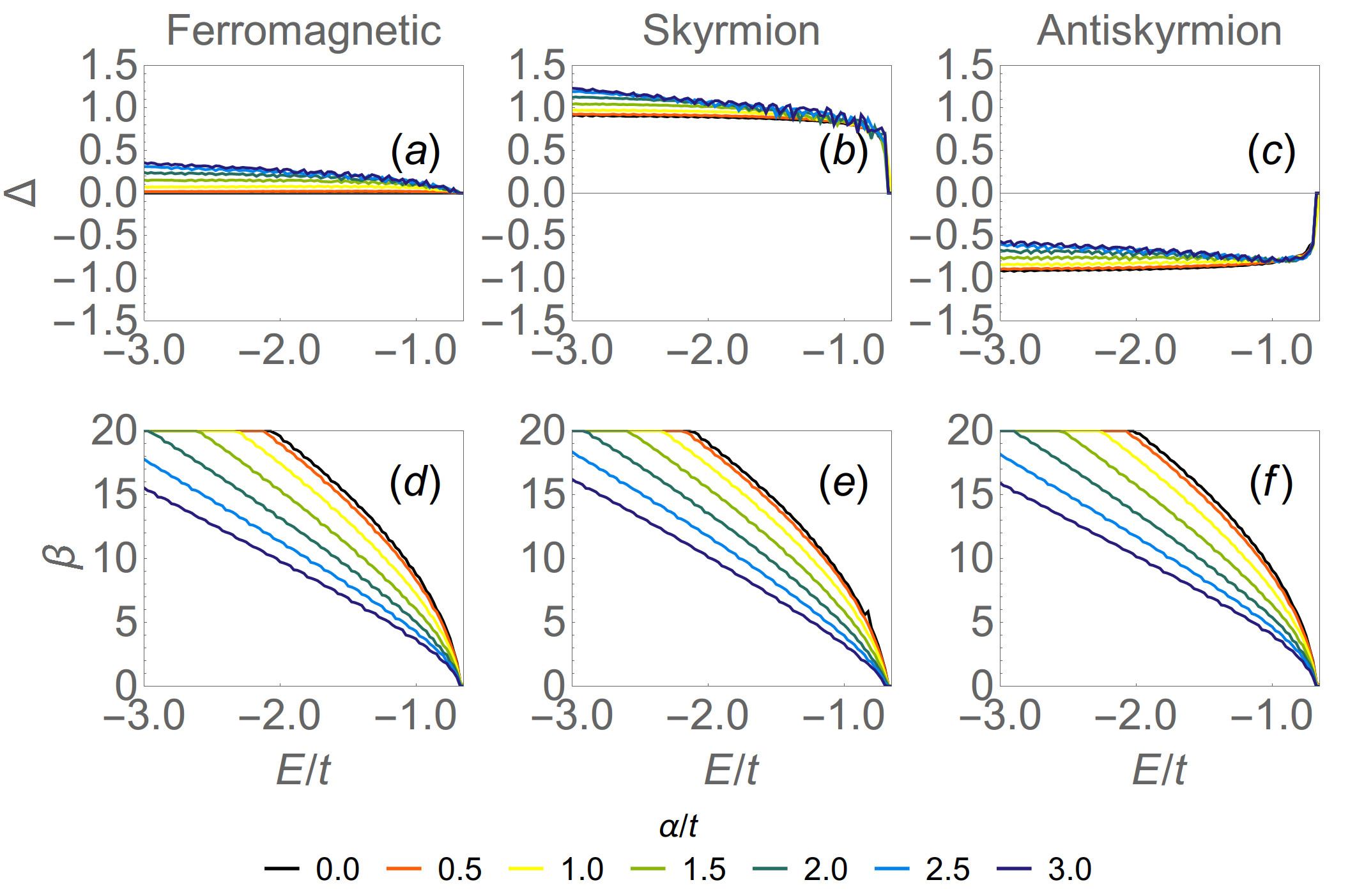}
\caption{(a), (b) and (c) show $\Delta$ as a function of $E$ for FM, skyrmion and antiskyrmion cases. (d), (e) and (f) reveal the energy dependence of $\beta$ for FM, skyrmion and antiskyrmion cases. Here we consider the parameter set (i). }
\label{fig:ABD}
\end{figure}

\begin{figure}[htp]
\centering
\includegraphics[width=0.5\textwidth]{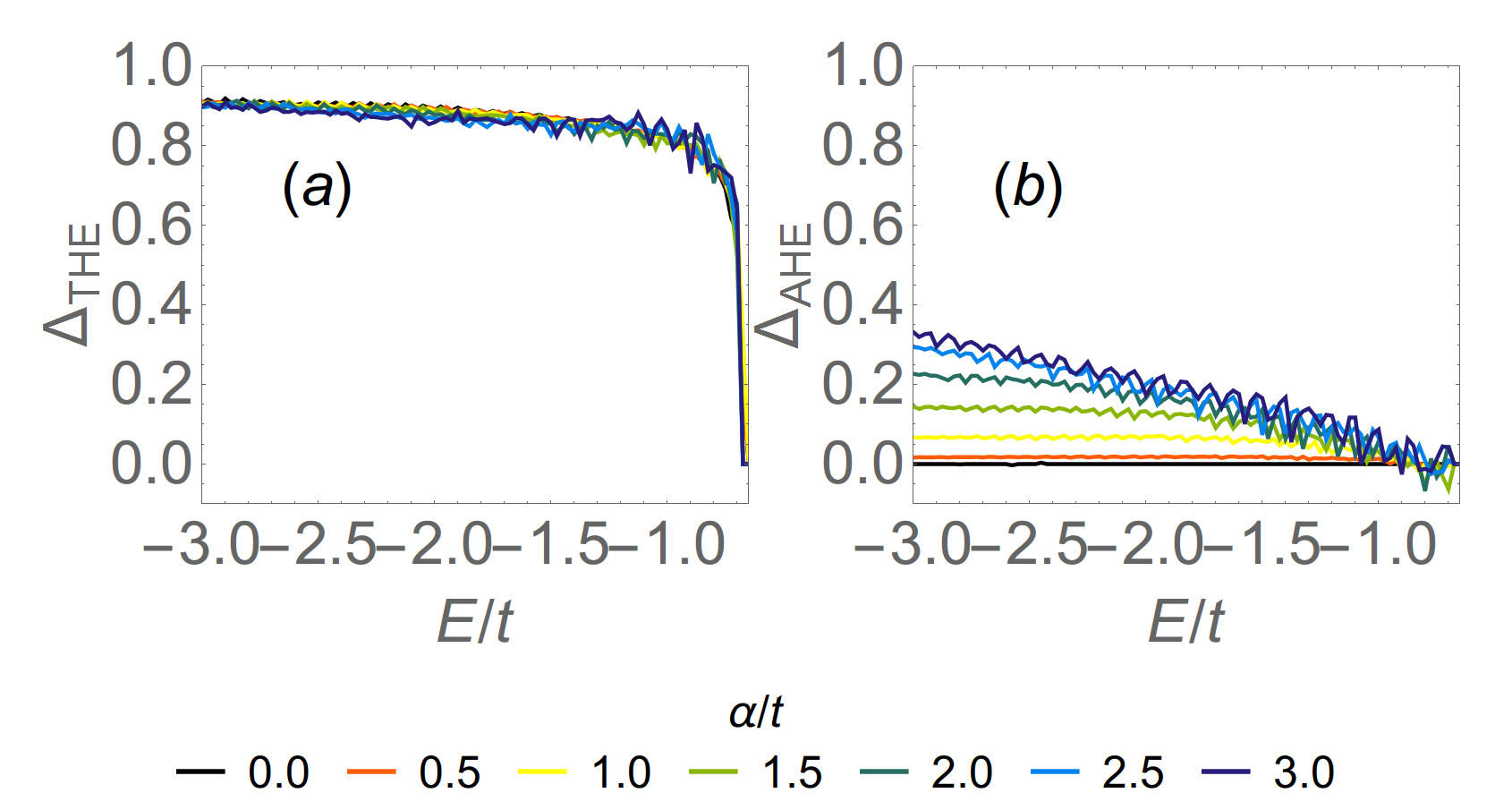}
\caption{(a) and (b) show the decomposition of $\Delta$ into the THE ($\Delta_{\textrm{THE}}$) and AHE part ($\Delta_{\textrm{AHE}}$), respectively. Here we choose the parameter set (i). }
\label{fig:DeltaDecompose}
\end{figure}

Eq.\;(\ref{eq:resistance_analytical}) and (\ref{eq:conductance_analytical}) are the basis for the analysis below. We can see that $\beta$ is related to the forward transmission and determines the longitudinal conductance $G_{xx}$ while $\Delta$ represents the asymmetry between the left and right scattering and determines the Hall conductance $G_{xy}$. In Fig.\;\ref{fig:ABD}, we demonstrate the behaviors of $\beta$ and $\Delta$ for the parameter set (i) as an example, from which we can understand the behaviors of AHE and THE.

Below we will analyze the behavior of $\Delta$ first (Fig.\;\ref{fig:ABD} (d) - (f)). In Fig.\;\ref{fig:ABD}(d), one can see that $\Delta$ increases when increasing the SOC parameter $\alpha$ in the FM case. From Fig.\;\ref{fig:ABD}(e) and (f), we find the value of $\Delta$ is much larger when there is a skyrmion or anti-skyrmion as compared to the FM case.
We may also consider a decomposition $ \Delta_{\textrm{Sk},n}=\Delta_{\textrm{AHE}}+ n \Delta_{\textrm{THE}}$ and the corresponding $\Delta_{\textrm{THE}}$ and $\Delta_{\textrm{AHE}}$ are plotted in Fig.\;\ref{fig:DeltaDecompose}(a) and (b), respectively.
One can see that all the curves for $\Delta_{\textrm{THE}}$ fall into one line and thus are independent of the SOC parameter $\alpha$, while $\Delta_{\textrm{AHE}}$ increases rapidly with $\alpha$. From Eq.\;(\ref{eq:conductance_analytical}), we expect that $G_{xy}$ exhibits a similar behavior as $\Delta$, which was indeed revealed in Fig.\;\ref{fig:THEenhancement}(c). Therefore, we conclude that SOC mainly increases the AHE contribution, but has little influence on the THE contribution to our four-band model in the clean limit.

Next, let us analyze the behavior of transmission $\beta$. For all three spin textures, the values of $\beta$ and the dependence of $\beta$ on SOC and the Fermi energy are quite similar, and thus we would not specify the magnetic texture for the discussion below. As expected, $\beta$ is decreasing when tuning Fermi energy to the valence band top (more insulating). In addition, we find a rapid decreasing of $\beta$ when increasing the SOC parameter $\alpha$ in Fig.\;\ref{fig:ABD}. This can be understood as the following. SOC tends to induce precession of electron spin and rotate it from the easy axis set by local magnetization. As a consequence, a strong scattering can be induced by the exchange coupling between electron spin and magnetic moments, and thus reduces transmission. It turns out that the reduction of transmission $\beta$ has a substantial influence on the behavior of Hall resistance $R_{xy}$. From Eq.\;\ref{eq:resistance_analytical}, we can see that $R_{xy}$ depends on the ratio between $\Delta$ and $\beta^2$. Therefore, although increasing SOC does not enhance $\Delta$, it reduces $\beta$, and thus increases $R_{xy}$ as shown in Fig.\;\ref{fig:THEenhancement}. (For a full figure of $R_{xy}$ for all SOC and spin textures, please see Appendix Fig.\;C1.) This analysis leads to the following conclusion for our model in the clean limit: (1) SOC does not have much influence on THE and (2) the behavior of $R_{\textrm{THE}}$ is mainly determined by the forward transmission $\beta$, rather than the asymmetric scattering $\Delta$.


\section{Analytical results of cross section}
In this section, we will provide more physical understanding on THE for our four-band model by analytically calculating the differential cross section of this system. We notice that topological surface states scattered by a magnetic skyrmion have been studied in Ref. \onlinecite{denisov2016electron, araki2017skyrmion}, while we focus on bulk QW states here. Due to the presence of spin-polarized background $S_z=\hat{z}$, we can treat $\bar{H}_0=H_0-m_0\sigma_z\otimes\tau_0$ as the unperturbed Hamiltonian, and take $\hat{V}(\bm{r}) =- (\bm{m}(\bm{r})\cdot\bm{\sigma}-m_0\sigma_z)\otimes\tau_0$ as the perturbation (scattering potential), where $\bm{m}(\bm{r})$ has been defined above in Eq. (\ref{eq:skyrmionmag}) and $m_0$ can be regarded as the exchange coupling strength between the conduction electron and local magnetic moment. The differential cross section of electron scattering is given by
\begin{equation}
\left(\frac{d\sigma(\phi)}{d\phi}\right)_{\alpha\beta} = |F_{\alpha\beta}(\bm{p},\bm{p'})|^2
\end{equation}
where $\phi$ is the scattering angle. $\psi_{{\bm p}'\alpha}$ and $\psi_{{\bm p}\beta}$ are eigenstates of $\bar{H}_0$ that describe the incident and scattered states respectively with ${\bm p}$ and ${\bm p}'$ is the associated momenta. $F_{\alpha\beta}=F_{\alpha\beta}(\bm{p},\bm{p'}) =  \langle\psi_{\bm{p'}\alpha}|\hat{V} + \hat{V}\hat{G}_0\hat{V}|\psi_{\bm{p}\beta}\rangle$ is the scattering amplitude up to the second order Born approximation, where $\hat{G}_0$ is the Green's function associated with unperturbed Hamiltonian $\bar{H}_0$. Asymmetric component of $\left(\frac{d\sigma(\phi)}{d\phi}\right)_{\alpha\beta}$ with respect to $\phi$, being responsible for the Hall response, arises from cross-terms between the first and second Born approximation.

\begin{figure}[htp]
\centering
\includegraphics[width=0.5\textwidth]{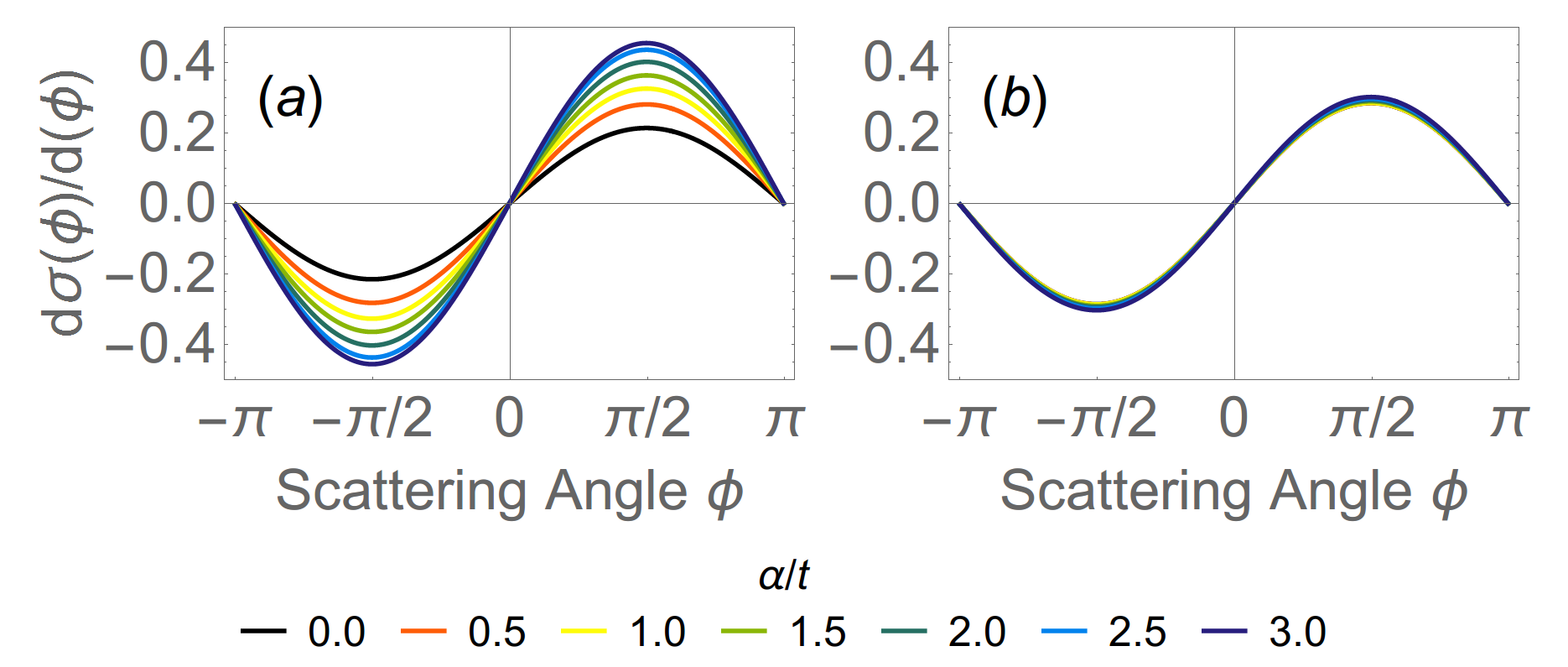}
\caption{Asymmetric part of the differential cross section as a function of scattering angle for (a) $m_0=5t$ and (b) $m_0=80t$. }
\label{fig:CrossSection}
\end{figure}

A major difficulty in this calculation is the computation of $\langle\psi_{\bm{p'}\alpha}|\hat{V}\hat{G}_0\hat{V}|\psi_{\bm{p}\beta}\rangle$ due to the Bessel function-like Green's function in 2D. Here we use the momentum representation so that $ \langle\psi_{\bm{p'}\alpha}|\hat{V}\hat{G}\hat{V}|\psi_{\bm{p}\beta}\rangle = \int d\bm{p}_1 \langle\alpha_{\bm{p'}}|V(\bm{p'}-\bm{p}_1)G(\bm{p}_1)V(\bm{p}_1-\bm{p})|\beta_{\bm{p}}\rangle$. For simplicity, we use a Bloch skyrmion configuration and let its polar angle in Eq. (\ref{eq:skyrmionmag}) be $\theta=\pi\exp(-r/a)$, where $a$ is the radius of the skyrmion. In the analytical calculations, Bessel functions of $qr$ will be used throughout the whole calculation, where ${\bm q}={\bm p}-{\bm p}'$. In the small angle scattering assumption, $qr\ll1$, so that we can expand the Bessel functions with Fourier series and keep the lowest order terms. In the current calculation, we are interested in the situation where the Fermi surface intersects only with the lowest electron band of $\bar{H}$. Direct calculation shows that the asymmetric part of the corresponding differential cross section is given by
\begin{eqnarray}
\left(\frac{d\sigma}{d\phi}\right)^A\propto a^4D(\varepsilon_F)m_0^3\left[\left(1-\cos\omega\right)\cos^4\frac{\omega}{2}\right.\nonumber\\
\left. + \left(1+\cos\omega\right)\sin^4\frac{\omega}{2}\right]\sin\phi
\end{eqnarray}
where $D(\varepsilon_F)$ is the density of states at the Fermi energy, and $\omega=\arccos [(Bp^2+M_0-m_0)/\varepsilon_F]$.
The cross-section as a function of scattering angle $\phi$ for different $m_0$ are shown in Fig.\;\ref{fig:CrossSection}.
With a small or intermediate exchange coupling strength $m_0$, we find that the asymmetric component of the differential cross section increases with the SOC parameter $\alpha$, and eventually saturate at large $\alpha$ limit, as shown in Fig.\;\ref{fig:CrossSection}(a). On the other hand, when $m_0$ is large, the influence of SOC parameter $\alpha$ becomes negligible due to the dominant role of exchange coupling in inducing THE in this regime, as shown in Fig.\;\ref{fig:CrossSection}(b), and our numerical results are consistent with the analytical result in this regime.


\section{Disorder effect}
\begin{figure}[htp]
\centering
\includegraphics[width=0.5\textwidth]{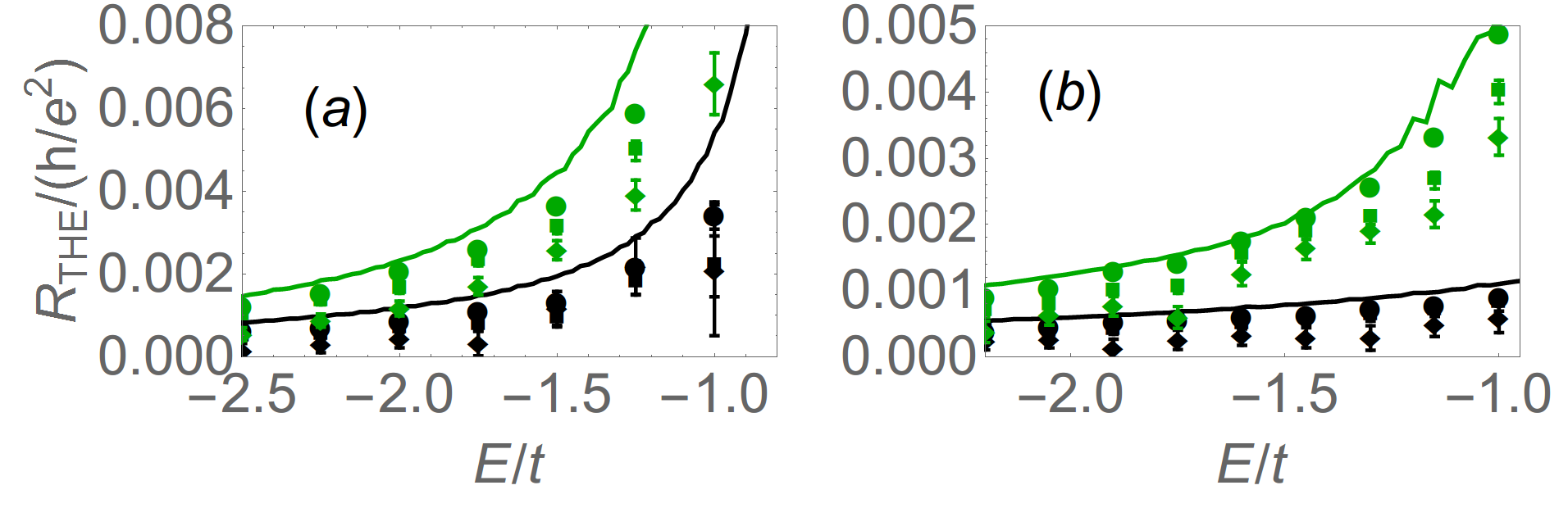}
\includegraphics[width=0.5\textwidth]{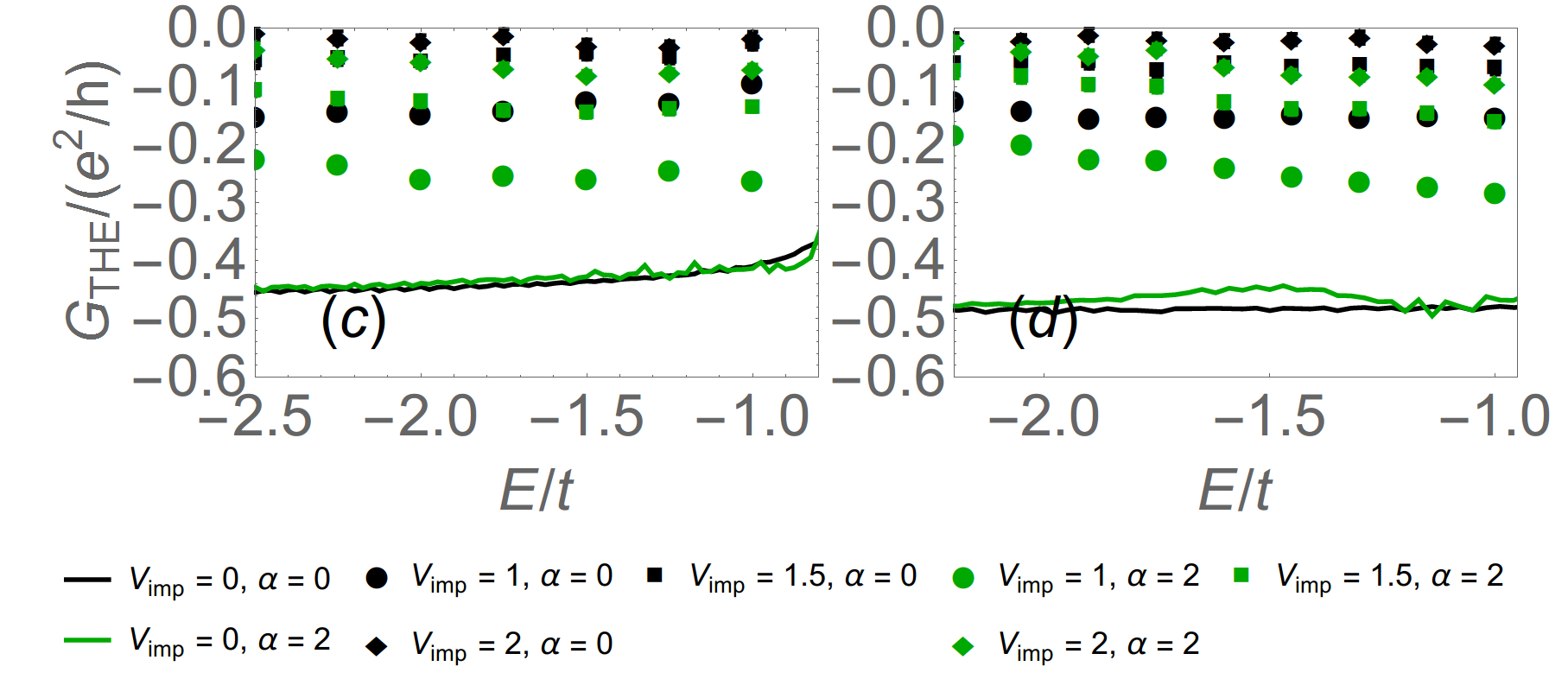}
\caption{(a) and (c) reveal $R_{\textrm{THE}}$ and $G_{\textrm{THE}}$ for the parameter set (i) under different disorder strength ($V_{\textrm{imp}}=0, 1, 1.5, 2$ in unit of $t$), respectively.
Green and black colors represent $\alpha/t = 2$ and $0$ cases, respectively.
(b) and (d) are the same as (a)(c), except that we choose the parameter set (ii). }
\label{fig:Disorder}
\end{figure}

\begin{figure}[htp]
\centering
\includegraphics[width=0.5\textwidth]{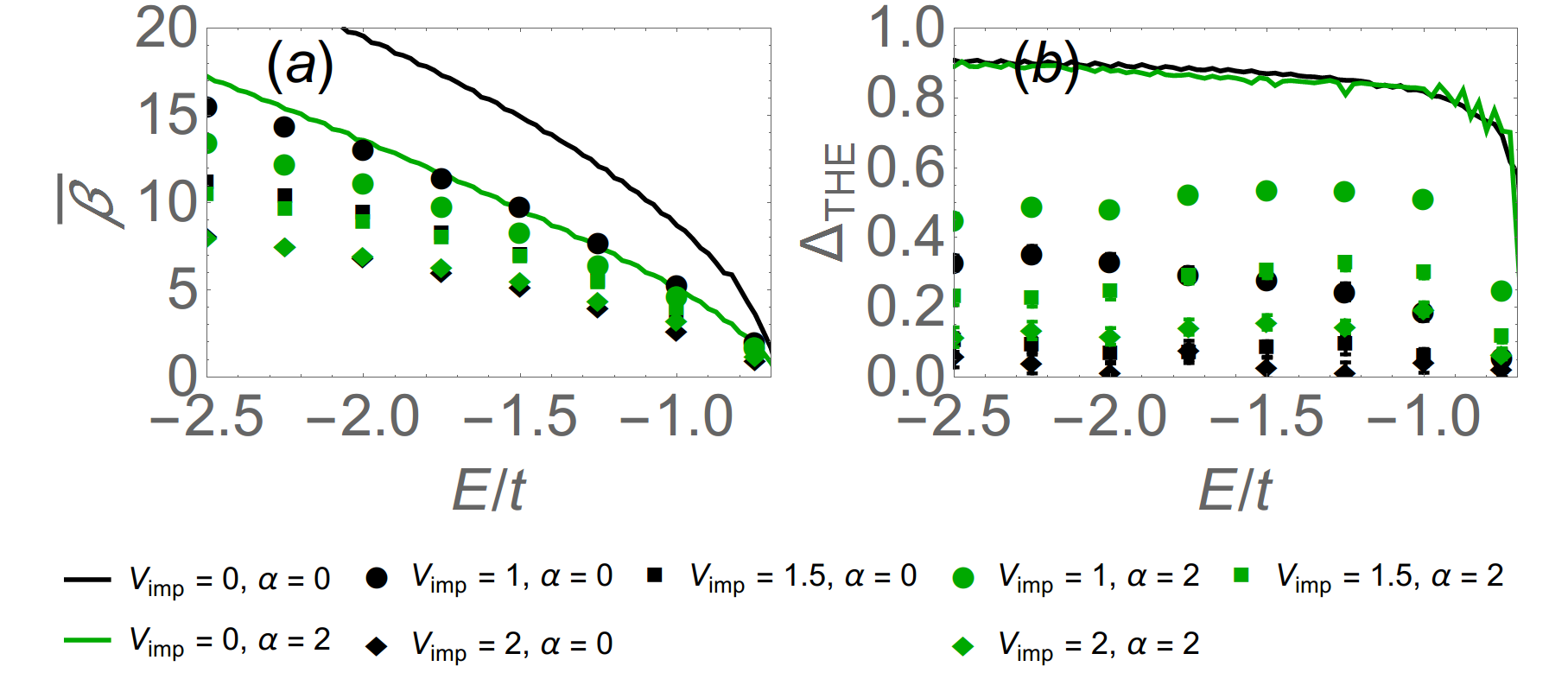}
\caption{(a) and (b) reveal $\bar{\beta}$ and $\Delta_{\textrm{THE}}$ as a function of $E$ in both disordered and clean limit. Here the colors represent different SOC strengths (green for $\alpha/t = 2$ and black for $\alpha/t=0$), and the line, circle, square and diamond represent $V_{\textrm{imp}}=0, 1, 1.5, 2$, respectively.    }
\label{fig:DisorderABD}
\end{figure}

We next examine the disorder effect on THE in MTI films, as shown in Fig.\;\ref{fig:Disorder}(a-d), which reveal new features compared to Fig.\;\ref{fig:THEenhancement} in the clean limit.
To consider the disorder effect, we introduce a spin-independent uniformly-distributed random on-site potential term $H_{\textrm{d}}=\sum_i \Psi^\dagger_i V_{\textrm{d},i}\Psi_i$ where
$V_{\textrm{d},i} = \textrm{Diag} (\mathcal{V}_{+}, \mathcal{V}_{+}, \mathcal{V}_{-}, \mathcal{V}_{-})$ and $\mathcal{V}_{\pm} \in \left[-V_{\textrm{imp}}/2, V_{\textrm{imp}}/2\right]$ with $0<V_{\textrm{imp}}\leq 2t$ chosen in our calculations.
All the calculations are performed with the disorder average over 160 samples. After such a sample average, the uncertainty (dictated by the error bars in Fig.\;\ref{fig:THEenhancement}) is much smaller than its mean value. We implement a similar decomposition of Hall conductance and resistance, as specified in Eq.\;(\ref{eq:Hall_skyrmion_decomposition}) and (\ref{eq:Hall_skyrmion_decomposition_conductance}) for each individual run, and the disorder-averaged Hall resistance ($R_{\textrm{THE}}$) and conductance ($G_{\textrm{THE}}$) from the THE contribution are revealed in Fig.\;\ref{fig:Disorder}(a, b) and (c, d) for two parameter sets (i) and (ii), respectively.
Here the circle, square and diamond label different disorder strength $V_{\textrm{imp}}=1, 1.5$ and $2$ in the unit of $t$, while the green and black colors are for different SOC strength ($\alpha/t = 0$ and $\alpha/t = 2$). The solid lines show the results in the clean limit for comparison.
With increasing the disorder strength, one can clearly see the decreasing of both Hall resistance $R_{\textrm{THE}}$ and conductance $G_{\textrm{THE}}$. A striking feature emerges in the Hall conductance $G_{\textrm{THE}}$ when increasing disorder strength. $G_{\textrm{THE}}$ is unchanged for different SOC strength in the clean limit, as seen by the coincidence between black and green solid lines in Fig.\;\ref{fig:Disorder} (c) and (d). In contrast, for intermediate or strong disorder strength, $G_{\textrm{THE}}$ at a large SOC $\alpha/t = 2$ can be much larger than that at zero SOC, as shown by the green and black markers in Fig.\;\ref{fig:Disorder}c and d. This suggests that SOC can stabilize the THE against disorder scattering.
We also analyze the disorder-averaged forward transmission $\beta$ and asymmetric scattering $\Delta$, shown in Fig.\;\ref{fig:DisorderABD}(a) and (b). Interestingly, we find that with increasing disorder scattering, forward transmission $\beta$, although being reduced for both SOC strengths, becomes comparable for $\alpha/t=0$ and $\alpha/t=2$ when $V_{\textrm{imp}}$ is increased above $1.5t$. This suggests that the mean-free path of electrons is determined by disorder, rather than SOC, at this disorder strength $V_{\textrm{imp}}/t=2$. In contrast, although the asymmetric scattering parameter $\Delta$ is reduced for both SOC strengths, its reduction is much slower when the SOC $\alpha$ is strong, which can be clearly seen by the fact that the green markers ($\alpha/t=2$) are above the black markers ($\alpha/t=0$) in Fig.\ref{fig:DisorderABD} (b). In contrast to the clean limit, in which SOC only enhances $R_{xy}$ but not $G_{xy}$ due to the reduction of transmission $\beta$, SOC mainly influences the asymmetric scattering $\Delta$ and thus will enhance both $R_{xy}$ and $G_{xy}$ in the disordered limit. Therefore, we conclude that the THE is stabilized by SOC in the disordered limit.


\section{Discussion and Conclusion}
In summary, we have studied the AHE and THE in an MTI model with skyrmion configuration and revealed how the magneto-transport behaviors in such systems are influenced by SOC, Fermi energy and disorder through numerical calculations and theoretical analysis. In particular, our calculations demonstrate the importance of disorder effect in determining the role of SOC in the THE. Given the recent experimental efforts in MTI systems \cite{yasuda2016geometric,liu2017dimensional}, our numerical and theoretical results will provide a physical understanding of these magneto-transport measurements and may stimulate further experimental studies. It should be pointed out that the SOC term we used here preserves inversion symmetry and thus is different from the Rashba SOC, which breaks inversion symmetry and responses for DM interaction. Including Rashba SOC in the calculation may bring new features and will require further studies, which is beyond the scope of the current work. In light of the importance of the interplay between disorder scattering and SOC, it will be important to develop a more analytical theory (such as the Boltzman equation and diagram expansion calculation \cite{nagaosa2010anomalous}) to take into account random scattering of multiple magnetic skyrmions or skyrmion lattice, SOC and disorder scattering, which will be another future direction. We would like to point out that although the current calculation is on a specific skyrmion texture, the presence of transverse scattering should exist for any spin textures, including random state, with net chirality\cite{hou2017thermally}. It will be interesting to generalize this work to the investigation of THE-AHE crossover in other chiral systems.

\section{Acknowledgement}
We acknowledge the discussion with C.Z. Chang and M.H.W. Chan. C.X.L and J.X.Z acknowledge support from the Office
of Naval Research (Grant No. N00014-15-1-2675 and renewal No. N00014-18-1-2793)
and the  U.S. Department of Energy (DOE), Office of Science, Basic Energy Sciences (BES) under award No. DE-SC0019064.
Work at UNH was supported by the U.S. Department of Energy (DOE), Office of Science, Basic Energy Sciences (BES) under Award No. DE-SC0016424.


\bibliography{main}
\end{document}